\documentclass[a4paper,11pt]{article}
\pdfoutput=1

\usepackage[margin=0.72in]{geometry}

\usepackage[utf8]{inputenc}
\usepackage[british]{babel}
\usepackage{csquotes} 
\usepackage[T1]{fontenc}

\usepackage{pifont}

\usepackage{amsmath,amssymb,amsthm}
\usepackage{mathtools}

\usepackage[
    labelfont=bf %
]{caption}
\usepackage{float}
\usepackage{subcaption}
\usepackage{tabularx}
\usepackage{tikz}
\usetikzlibrary{decorations.markings,arrows,patterns,patterns.meta,shapes}
\usetikzlibrary{positioning}

\usepackage{pgfplots}
\pgfplotsset{compat=1.18}
\pgfdeclareplotmark{donut}{
    \pgfseteorule
    \pgfpathellipse{\pgfpoint{0cm}{0cm}}
             {\pgfpoint{3pt}{0cm}}
             {\pgfpoint{0cm}{1pt}}
     \pgfpathellipse{\pgfpoint{0cm}{0cm}}
             {\pgfpoint{4pt}{0cm}}
             {\pgfpoint{0cm}{2pt}}
    \pgfusepath{fill}
}
\pgfdeclareplotmark{donut*}{
     \pgfpathellipse{\pgfpoint{0cm}{0cm}}
             {\pgfpoint{4pt}{0cm}}
             {\pgfpoint{0cm}{2pt}}
    \pgfusepath{fill}
}

\usepackage[outline]{contour}
\contourlength{0.2em}

\usepackage[compat=0.6]{yquant}
\useyquantlanguage{groups}

\usepackage{url}

\usepackage{mathrsfs}
\usepackage{dsfont}
\usepackage{setspace}
\usepackage{verbatim}
\usepackage[export]{adjustbox}
\usepackage{tqft}
\usepackage[inline]{enumitem}

\usepackage{quantikz}
\usepackage{graphicx}

\usepackage{bbm}

\usepackage{xcolor}
\usepackage{bm}
\definecolor{darkblue}{RGB}{0,0,128}
\definecolor{darkgreen}{RGB}{0,150,0}

\definecolor{dark-red}{rgb}{0.4,0.15,0.15}
\definecolor{dark-blue}{rgb}{0.15,0.15,0.4}
\definecolor{dark-green}{rgb}{0.15,0.4,0.15}
\definecolor{medium-blue}{rgb}{0,0,0.5}

\definecolor{drawing-blue}{HTML}{c1ecff}
\definecolor{drawing-yellow}{HTML}{f6dab5}
\definecolor{drawing-red}{HTML}{e5bdd2}

\usepackage[%
	backend=biber,
	style=alphabetic,
	url=false,
	isbn=true,
	maxbibnames=10,
	backref, 
]{biblatex}
\usepackage[pdfusetitle]{hyperref}
\hypersetup{
    breaklinks,
    colorlinks,
	linkcolor={dark-blue},
	citecolor={dark-green},
	urlcolor={medium-blue},
    filecolor=red,
}

\usepackage[nameinlink,capitalise]{cleveref}
\crefname{figure}{Figure}{Figures}

\crefname{conjecture}{Conjecture}{Conjectures}
\Crefname{conjecture}{Conjecture}{Conjectures}

\newtheorem*{theorem*}{Theorem}

\usepackage{thmtools}
\usepackage{thm-restate}

\usepackage{braket}
\usepackage{dsfont}
\usepackage[normalem]{ulem}
\usepackage{algorithm}
\usepackage{algpseudocode}
\usepackage{newfloat}


\usepackage{stmaryrd}

\newcommand{\sansserif}[1]{%
  \ifmmode
  \mathsf{#1}%
  \else
   \textsf{#1}%
  \fi
}

\usepackage[mathscr]{euscript}

\usepackage{mathtools}

\DeclarePairedDelimiterXPP\bigo[1]{O}{(}{)}{}{#1}
\DeclarePairedDelimiterXPP\littleo[1]{o}{(}{)}{}{#1}
\DeclarePairedDelimiterXPP\bigomega[1]{$\Omega$}{(}{)}{}{#1}
\DeclarePairedDelimiterXPP\bigtheta[1]{$\Theta$}{(}{)}{}{#1}

\graphicspath{{../}{./}{./tikz}{figs/}}

\renewcommand{\sec}[1]{\hyperref[sec:#1]{Section~\ref*{sec:#1}}}
\newcommand{\app}[1]{\hyperref[app:#1]{Appendix~\ref*{app:#1}}}
\newcommand{\ssec}[1]{\hyperref[ssec:#1]{Subsection~\ref*{ssec:#1}}}
\newcommand{\fig}[1]{\hyperref[fig:#1]{Figure~\ref*{fig:#1}}}
\newcommand{\tab}[1]{\hyperref[tab:#1]{Table~\ref*{tab:#1}}}
\newcommand{\lemm}[1]{\hyperref[lemm:#1]{Lemma~\ref*{lemm:#1}}}
\newcommand{\propos}[1]{\hyperref[propos:#1]{Proposition~\ref*{propos:#1}}}
\newcommand{\thm}[1]{\hyperref[thm:#1]{Theorem~\ref*{thm:#1}}}
\newcommand{\alg}[1]{\hyperref[alg:#1]{Algorithm~\ref*{alg:#1}}}
\newcommand{\defn}[1]{\hyperref[defn:#1]{Definition~\ref*{defn:#1}}}

\newcommand{\be}{\begin{equation}}
\newcommand{\ee}{\end{equation}}
\newcommand{\bea}{\begin{eqnarray}}
\newcommand{\eea}{\end{eqnarray}}

\newcommand{\bem}{\begin{multline}}
\newcommand{\eem}{\end{multline}}

\usepackage{array}
\usepackage{booktabs}
\usepackage{makecell}
\usepackage{diagbox}
\usepackage{multicol}
\usepackage{multirow}

\usepackage{siunitx}

\newcolumntype{?}{!{\vrule width 1pt}}

\usepackage{authblk}

\title{Forced Gap Post-Selection for\\Quantum LDPC Codes and their Operations}
\author[1,2,3]{Adam Wills\thanks{Email: a\_wills@mit.edu}}
\author[1]{Theodore J. Yoder}
\author[3,4]{Isaac Chuang}

\affil[1]{IBM Quantum\vspace*{0.2cm}}
\affil[2]{Center for Theoretical Physics — a Leinweber Institute\linebreak Massachusetts Institute of Technology, Cambridge, MA\vspace*{0.2cm}}

\affil[3]{The NSF AI Institute for \linebreak Artificial Intelligence and Fundamental Interactions\vspace*{0.2cm}}
\affil[4]{Department of Electrical Engineering and Computer Science\linebreak Massachusetts Institute of Technology}

\date{\today}

\begin{document}
\maketitle

\begin{abstract}
    We develop a simple and general post-selection strategy for high-rate quantum codes that is transferable across decoders. After an initial baseline run, the decoder is re-run once per logical observable, and forced in these latter runs to provide a solution where the given observable has the complementary outcome. Shots are rejected that find logically complementary solutions with similar likelihoods compared to the baseline. Using the Relay-BP decoder, we benchmark the strategy on the $72$-qubit and $144$-qubit bivariate bicycle codes, as well as surgery gadgets for the latter. In comparison to previous post-selection strategies, our results offer an improved logical error rate by over a factor of $4$ on the same circuit and physical error rate, and at the same rate of post-selection. Our strategies are also lightweight, relying only on FPGA-friendly belief propagation, whereas the previous best used repeated rounds of a high-latency BP-OSD decoder.
\end{abstract}

\begin{multicols}{2}

\paragraph*{Introduction}

Is it possible that a decoder for a quantum code could return not only a most-likely correction, but also some quantification of the likelihood of a logical error having occurred? If so, it would be possible to \textit{post-select} (reject) suspicious decoding instances, in exchange for a lower logical error rate on surviving shots. For the surface code~\cite{kitaev2003fault}, this is known to be readily possible by computing a \textit{complementary gap}~\cite{hutter2014efficient,bombin2024fault,gidney2024magic,gidney2025yoked}, a confidence measure comparing the log-likelihoods of the most-likely correction and a logically complementary one. High-rate quantum low-density parity check (LDPC) codes~\cite{breuckmann2021quantum} can offer a greatly improved space overhead over the surface code, although one (of several) practical challenges for LDPC codes is that post-selection strategies based on complementary gaps do not directly apply to them. Despite this, recent works~\cite{lee2025efficient,xie2026simple} have initiated the development of such strategies.

In this work, we develop a new idea for post-selection applicable to high-rate quantum LDPC codes. At its core, this idea is a simple extension of the complementary gap strategy for surface codes to situations where many logical observables are present. We benchmark our post-selection strategy using the Relay-BP decoder~\cite{muller2025improved} on bivariate bicycle (BB) codes, namely idling circuits for the $72$-qubit code and the $144$-qubit ``gross'' code~\cite{bravyi2024high}, as well as logical operations for the latter constructed via LDPC code surgery in~\cite{yoder2025tourgrossmodularquantum}.

\paragraph*{Problem Setup}

In the surface code, the \textit{complementary gap} for a given logical observable is the log-likelihood difference between the most-likely correction and the most-likely correction consistent with the complementary logical outcome for that observable. A small complementary gap indicates little difference in likelihoods between these two outcomes, implying a high probability of a logical error. It is computed as follows. Treating $X$ and $Z$ errors separately, one first performs a standard decode with a matching decoder to obtain the most-likely correction. Then, the decoding instance is re-run while being forced to produce a correction with the complementary logical outcome; for the $X$ observable, for example, the sum of the two corrections is required to be a logical $X$ operator. In the surface code, this modified decoding instance is still matchable and so may be efficiently solved. The log-likelihood difference between the two solutions gives the complementary gap.

The complementary gap technique does not extend immediately to general quantum LDPC codes for two important reasons. First, general quantum LDPC codes cannot be decoded by matching decoders. Second, circuits corresponding to high-rate quantum LDPC codes have many observables, rendering na\"{i}ve comparative decoding of complementary logical classes infeasible. For example, a code with $k$ logical qubits would have $4^k$ distinct logical classes, even in an idling circuit. Even for relatively simple quantum LDPC codes such as the $72$ and $144$-qubit BB codes, one has $k = 12$, and so attempting to decode a most-likely correction in each logical class would be infeasible. In this work, we develop an efficient analogue that overcomes both obstacles, extending the complementary gap idea to general quantum LDPC codes with many logical observables.

\paragraph*{Methods}\mbox{}\\

\noindent{\it Decoding setup.} We describe decoding problems using a parity-check matrix $H \in \mathbb{F}_2^{M \times N}$, action matrix $A \in \mathbb{F}_2^{K \times N}$, and prior probability vector $p\in[0,1/2)^N$.\footnote{Briefly, matrix entry $H_{ij}$ indicates whether detector \cite{gidney2021stim} $i$ detects fault $j$ occurring alone, $A_{ij}$ indicates whether observable $i$ is flipped by fault $j$ alone, and $p_j$ is the probability fault $j$ occurs (with the assumption that each fault occurs independently). See~\cite{ott2025decision} for a thorough description of this formulation of the decoding problem, which captures many cases of interest including those studied here.} This represents a decoding problem with $N$ possible errors, $M$ detectors, and $K$ logical observables.\footnote{For example, we have $K = 24$ observables for an idling code with $k=12$ logical qubits, since there are $12$ $X$ observables and $12$ $Z$ observables.} Given a decoding problem, a decoding instance is specified by an observed syndrome $\sigma \in \mathbb{F}_2^M$ and an unobserved fault configuration $f\in\mathbb{F}_2^N$ satisfying $H\cdot f=\sigma$. 

The decoder's job is to find a correction $e$ such that $H\cdot e=\sigma$ with a successful decode characterised by $A\cdot e=A\cdot f$, meaning the logical observables corrupted by faults $f$ have been corrected by $e$. A reasonable way to frequently decode successfully is to choose a correction that has a large likelihood $\mathbb{P}[e]=\prod_{j=1}^N(1-p_j)^{1-e_j}p_j^{e_j}$ of having occurred given the syndrome. Decoders may attempt to do this heuristically \cite{mackay2004sparse,liu2019neural,muller2025improved} or exactly \cite{ott2025decision,beni2025tesseract}.

\noindent{\it Motivating scenario.} We motivate our post-selection strategy by considering what is possible if we had the ability to perform maximum-likelihood decoding (MLD), including the ability to calculate the likelihood of the most-likely logical error class. Though what we describe in this thought experiment is computationally infeasible in practice, it motivates our actual proposal which replaces certain steps with more practical computations approximating the MLD version.

As a quick review, MLD maximises the likelihood of a successful correction by returning a correction from the most-likely logical class $\lambda^*=\mathrm{argmax}_{\lambda\in\mathbb{F}_2^K}\mathbb{P}_L[\lambda]$, where $\mathbb{P}_L[\lambda]:=\sum_{e\in\mathbb{F}_2^N}\mathbb{P}[e]\hspace{2pt}\mathbbm{1}\hspace{-2pt}\left[H\cdot e=\sigma\right]\hspace{2pt}\mathbbm{1}\hspace{-2pt}\left[A\cdot e=\lambda\right]$ is the probability of logical class $\lambda$ and syndrome $\sigma$. We assume MLD returns $\mathbb{P}_L[\lambda^*]$ as well as a correction.

To define a post-selection strategy, our goal is to calculate 
\begin{equation}
\mathrm{Exact\;Gap}:=\log\mathbb{P}_L[\lambda^*]-\log\mathbb{P}_L[\lambda^{**}],
\end{equation}
where $\lambda^{**}$ is the second most-likely logical class. The Exact Gap is a natural generalisation of the complementary gap to many logical observables: rather than comparing the most-likely correction with a single logically complementary one, it compares the two most-likely logical classes overall. It is a reasonable measure of decoding confidence; if it is small, then the most-likely logical class and the second most-likely have similar likelihoods, and our confidence in a successful decode should be low. For instance, a post-selection threshold $T$ can be imposed and decoding instances thrown out if their Exact Gap is below $T$. To calculate the Exact Gap, we can use MLD directly to get the first term $\log\mathbb{P}_L[\lambda^*]$ and, as we argue now, $K$ instances of MLD on modified problems to get the second term $\log\mathbb{P}_L[\lambda^{**}]$.

The second most-likely logical class $\lambda^{**}$ must differ from the most-likely class $\lambda^*$ in at least one of its $K$ bits. If we write $\lambda^{*i}$ to indicate the most-likely logical class of those classes differing from $\lambda^*$ in the $i$'th bit, then $\mathbb{P}_L[\lambda^{**}]=\max_{i=1,\dots,K}\mathbb{P}_L[\lambda^{*i}]$, reducing the calculation of $\mathbb{P}_L[\lambda^{**}]$ to the calculation of each of $\mathbb{P}_L[\lambda^{*i}]$. 

To calculate $\mathbb{P}_L[\lambda^{*i}]$, set up a modified decoding problem in which the $i$'th row of $A$ is appended to the bottom of the parity-check matrix $H$ to get a modified matrix $H^{(i)}$. The syndrome is also modified by appending a bit $1-\lambda^{*}_i$ to $\sigma$ to get vector $\sigma^{(i)}$. MLD finds a correction $e^{(i)}$ with some logical class $L^{(i)}=A\cdot e^{(i)}$ and consistent with the syndrome, $H^{(i)}\cdot e^{(i)}=\sigma^{(i)}$. By construction of $H^{(i)}$, this latter vector equality enforces exactly $H\cdot e^{(i)}=\sigma$ and $L^{(i)}_j=1-\lambda^{*}_i$ (from the last row of $H^{(i)}$). Thus, by running MLD on the modified decoding instance $(H^{(i)},A,p,\sigma^{(i)})$, we obtain a correction from $L^{(i)}=\lambda^{*i}$ and also the according value of $\mathbb{P}_L[\lambda^{*i}]$. Doing this for all $i=1,\dots,K$, we complete the calculation of Exact Gap. Notably, this computation required much fewer than the na\"{i}ve $2^K$ decoding invocations: just 1 invocation on the baseline problem plus $K$ invocations on modified problems.

\noindent{\it Forced gap post-selection.} In practice, we do not have access to MLD or to the exact calculation of the logical class likelihoods $\mathbb{P}_L[\lambda]$. Instead, for our post-selection strategy, we assume a decoder that operates on arbitrary $(H,A,p,\sigma)$ decoding instances but also allow the decoder to ``give up" when it is unable to return any valid correction.\footnote{Functionally, an un-converged decode is a heralded failure. If the context of the error-corrected computation still demands some correction consistent with the syndrome, $H\cdot e=\sigma$ can be solved with linear algebra, assuming either that $H$ has full row rank, or that the real noise that generated the syndrome is not inconsistent with the decoding problem model. In practice, the logical information in the module is considered lost, and the codestate may be returned to some fixed state such as the all-zeros state.} Such ``un-converged runs'' are part of the nature of purely iterative decoders like BP or Relay-BP \cite{muller2025improved}, and so we accommodate that potential behavior within our general post-selection framework. 

\begin{figure*}[t]
\centering
\begin{tikzpicture}[
  box/.style  = {draw, rounded corners=4pt, align=center,
                 font=\small, inner sep=5pt, minimum height=0.82cm},
  arr/.style  = {->, >=stealth, thick},
  sarr/.style = {->, >=stealth, semithick},
]

\fill[drawing-blue!50, rounded corners=5pt]
     (-0.25,-1.70) rectangle (4.60,1.55);

\node[box, fill=drawing-blue, draw=black!40, minimum width=1.0cm] (syn)
     at (0.65,0.20) {$\sigma$};
\node[box, fill=drawing-blue, draw=black!40, minimum width=1.8cm] (dec)
     at (3.10,0.20) {Baseline\\Decoder};
\draw[arr] (syn) -- (dec);

\node[box, fill=red!12, draw=red!50, minimum width=1.7cm] (p1reject)
     at (3.10,-1.10) {\textbf{Reject}};
\draw[arr, red!70!black]
     (dec.south) -- (p1reject.north)
     node[midway, right, font=\scriptsize] {erasure};

\node[font=\scriptsize\bfseries, text=black!55]
     at (2.175,1.32) {\textsc{Phase 1 --- Baseline Run}};

\fill[drawing-yellow!55, rounded corners=5pt]
     (4.825,-2.15) rectangle (11.225,1.55);

\node[box, fill=drawing-yellow!80, draw=black!25,
      minimum width=2.6cm, font=\footnotesize,
      text=black!75] (mod)
     at (6.575,0.0)
     {$H^{(i)}\leftarrow\!\begin{bmatrix}H\\ A_i\end{bmatrix}$\\[3pt]
      $\sigma^{(i)}\leftarrow\!\begin{bmatrix}\sigma\\ 1\oplus L^{(0)}_i\end{bmatrix}$};

\draw[arr] (dec.east) -- (mod.west)
     node[midway, above, font=\scriptsize, text=black!60]
     {$\,e^{(0)}\!,L^{(0)}$};

\node[box, fill=drawing-yellow, draw=black!45, minimum width=2.3cm] (fr1)
     at (9.725, 0.75) {Forced Run $1$};
\node[box, fill=drawing-yellow, draw=black!45, minimum width=2.3cm] (fr2)
     at (9.725,-0.05) {Forced Run $2$};
\node[font=\large] at (9.725,-0.72) {$\vdots$};
\node[box, fill=drawing-yellow, draw=black!45, minimum width=2.3cm] (frK)
     at (9.725,-1.52) {Forced Run $K$};

\draw[arr] (mod.east) to[out=15,in=180]  (fr1.west);
\draw[arr] (mod.east) to[out= 0,in=180]  (fr2.west);
\draw[arr] (mod.east) to[out=-25,in=180] (frK.west);

\node[font=\scriptsize\bfseries, text=black!55]
     at (8.025,1.32) {\textsc{Phase 2 --- $K$ Forced Runs}};

\fill[drawing-red!45, rounded corners=5pt]
     (11.45,-2.15) rectangle (16.30,1.85);

\node[box, fill=white, draw=black!20, minimum width=4.0cm,
      font=\footnotesize] (rank)
     at (13.875,0.80)
     {Pool all $(e^{(i)},\,L^{(i)})$;\\ Find $\lambda^{(0)},\,\lambda^{(1)}$};

\node[box, fill=drawing-red!75, draw=black!35,
      minimum width=4.0cm, font=\footnotesize] (gap)
     at (13.875,-0.30)
     {Gap $=\log\mathbb{P}[\lambda^{(0)}]-\log\mathbb{P}[\lambda^{(1)}]$};

\draw[arr] (rank) -- (gap);

\node[box, fill=red!12, draw=red!50, minimum width=1.7cm] (reject)
     at (12.55,-1.55) {\textbf{Reject}};
\node[box, fill=green!15, draw=green!55, minimum width=1.7cm] (accept)
     at (15.20,-1.55) {\textbf{Accept}};

\draw[arr, red!70!black]
     (gap.south) to[out=228,in=90]
     node[left, font=\scriptsize] {$<T~~~~$} ([xshift=0ex]reject.north);
\draw[arr, green!60!black]
     (gap.south) to[out=312,in=90]
     node[right, font=\scriptsize] {$~~~~~\ge\, T$} (accept.north);

\draw[sarr] (fr1.east) to[out=0,in=180] (rank.west);
\draw[sarr] (fr2.east) to[out=0,in=180] (rank.west);
\draw[sarr] (frK.east) to[out=0,in=180] (rank.west);

\node[font=\scriptsize\bfseries, text=black!55]
     at (13.875,1.62) {\textsc{Phase 3 --- Gap \& Decision}};

\end{tikzpicture}
\caption{%
  Overview of the Forced Gap post-selection strategy.
  {Phase~1 (blue):} the decoder is run on syndrome~$\sigma$ to produce a
  baseline correction~$e^{(0)}$ with logical class~$L^{(0)} = A\cdot e^{(0)}$.
  If the decoder declares erasure the instance is immediately rejected.
  {Phase~2 (amber):} for each of the $K$ logical observables ($i=1,\dots,K$),
  a forced run solves the modified problem $(H^{(i)},\sigma^{(i)})$, where an
  extra row appended to the parity-check matrix constrains the $i$-th observable to
  the complementary outcome $1\oplus L^{(0)}_i$.
  {Phase~3 (pink):} the corrections from all runs are pooled; the two
  most-likely distinct logical classes $\lambda^{(0)},\lambda^{(1)}$ are identified
  and their log-likelihood difference, the Gap, is computed.  Instances with
  Gap~$<T$ are rejected (post-selected); the remainder are accepted.  The total
  decoder invocation count is $K+1$, scaling as $\mathcal{O}(K)$ rather than the
  na\"{i}ve $2^K$. In addition, the latter $K$ invocations may be performed in parallel.%
}
\label{fig:method_overview}
\end{figure*}
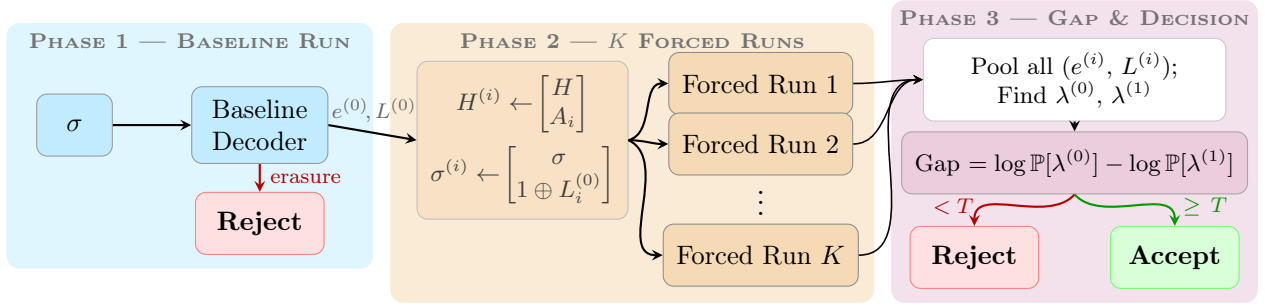

Our method for developing a post-selection strategy (illustrated in Fig.~\ref{fig:method_overview}) is as follows. Given a decoding instance, we begin by running the chosen decoder as normal in what is called the \textit{baseline run}. Provided the baseline run converged to some valid solution $e^{(0)} \in \mathbb{F}_2^N$, its logical class is calculated as $L^{(0)} \coloneq A\cdot e^{(0)}$. Then, $K$ further runs of the decoder take place. These are called the $K$ \textit{forced runs}. In the $i$'th forced run, the decoder is forced to find a correction $e^{(i)}\in \mathbb{F}_2^N$ where the $i$'th observable has the complementary logical outcome compared to $L$, or, in other words, the $i$'th bit of $L^{(i)} \coloneq A\cdot e^{(i)}$ differs from the $i$'th bit of $L^{(0)}$. This is done by running the decoder on the modified decoding instance $(H^{(i)},A,p,\sigma^{(i)})$ defined previously.

Across the baseline and all forced runs that converged on some solution, every candidate correction has its logical class recorded, but only some of these are distinct.
The set of distinct logical classes found is denoted by
\begin{equation}
    \Lambda = \{\lambda^{(0)}, \lambda^{(1)}, \ldots\}, \text{ where } \lambda^{(j)} \in \mathbb{F}_2^K,
\end{equation}
and each of these classes is associated with likelihoods 
\begin{equation}
\mathbb{P}\left[\lambda^{(j)}\right]:=\max_{i:\, L^{(i)} = \lambda^{(j)}}\left\{\,\mathbb{P}[e^{(i)}] \right\}
\end{equation}
(the likelihood of the most-likely correction found with that logical class). We order the set $\Lambda$ by decreasing likelihood so that $\mathbb{P}[\lambda^{(0)}]\ge\mathbb{P}[\lambda^{(1)}]\ge\dots$.

We then define the \textit{gap} to be the difference in log-likelihoods of the two most-likely logical classes, that is,
\begin{equation}
    \text{Gap} \coloneq \log\;\mathbb{P}\left[\lambda^{(0)}\right] - \log\;\mathbb{P}\left[\lambda^{(1)}\right].
\end{equation}
The Gap is our practical analogue of the complementary gap: it approximates the Exact Gap by using the likelihoods of specific decoder-found corrections rather than exact logical-class probabilities. In the case that all forced runs were un-converged, so $\Lambda$ is a set of size one and $\lambda^{(1)}$ does not exist, we choose to set $\text{Gap} = \infty$. If the baseline run was itself un-converged, $\Lambda$ is empty and we choose to set $\text{Gap}=0$. In this case, the decoding instance is termed an ``erasure''.\footnote{Note that it is also possible to have $\text{Gap} = 0$ for non-erasure shots, if two logically distinct solutions are found with exactly the same likelihood, and indeed, such instances are found in testing.}

We create a family of post-selection strategies, named the \textbf{forced gap} strategies, by picking a threshold $T \in \mathbb{R}_{\geq 0}$. Decoding instances are then rejected if the gap was found to be less than $T$. Succinctly,
\begin{multline}
    \text{Forced Gap }(T) = \left[\text{Reject if Gap}\; < \;T\right].
\end{multline}
There is no post-selection when $T=0$, and erasure events are considered logical errors. In contrast, erasures are always rejected when $T$ is positive, since then $\text{Gap} = 0$ by convention. Moreover, non-erasure instances where all forced runs were un-converged are always accepted, since then $\text{Gap} = \infty$ by convention.

At first, running $K$ forced runs might seem computationally daunting, although note that the forced runs may be run in parallel. Additionally, the total amount of computation scales linearly with $K$, giving the same amount of computation if the same logical information were encoded in (multiple) low rate codes, like surface codes.

Before moving on, we note that if a single run of the decoder finds multiple valid corrections per decoding instance, possibly logically distinct from one another, these can all be used to create the set of logical classes $\Lambda$, improving the forced gap strategy. Relay-BP, for example, explicitly has a parameter denoting the number of solutions it should find before terminating a single decoding instance (see Appendix~\ref{sec:parameter_choices} for brief review of Relay-BP), and we indeed use all solutions in our simulations.


\begin{figure*}[t]
  \centering

  \begin{subfigure}[t]{0.48\textwidth}
    \centering
    \includegraphics[width=\linewidth]{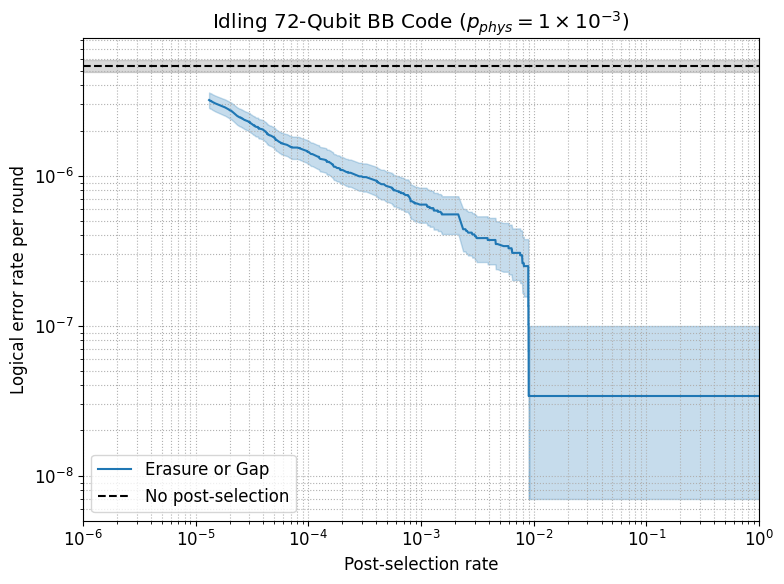}
    \caption{}
  \end{subfigure}
  \hfill
  \begin{subfigure}[t]{0.48\textwidth}
    \centering
    \includegraphics[width=\linewidth]{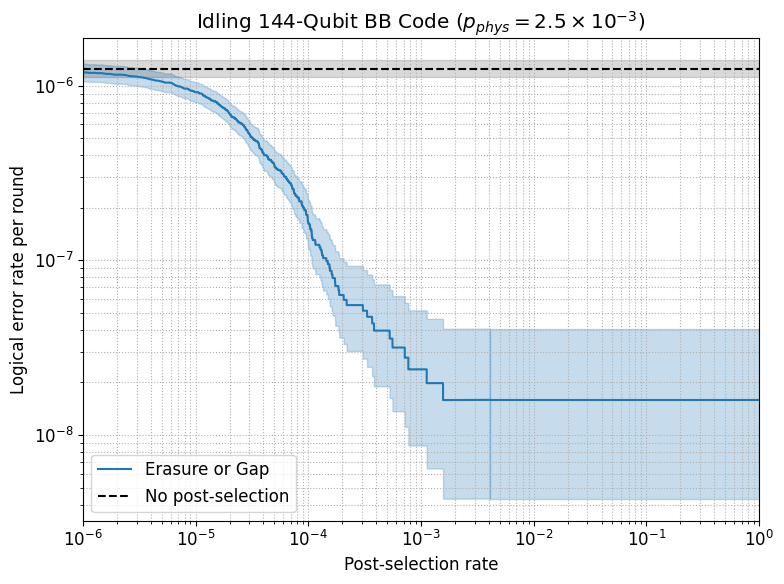}
    \caption{}
  \end{subfigure}

  \caption{Simulated error rates per round for the idling $[[72,12,6]]$ code, and the $[[144,12,12]]$ code under post-selection, at physical noise $p = 10^{-3}$ and $2.5\times 10^{-3}$, respectively. The $x$-axis shows ``Post-selection rate'', that is, the fraction of shots that are rejected, where a curve is generated by varying the threshold $T$. Both codes are simulated for $6$ and $12$ rounds of syndrome extraction, respectively, and the logical error rate is then normalised per round (we do not normalise by the number of logical qubits). The shaded regions are $95\%$ confidence intervals. The flat horizontal sections on the right of both graphs indicate the regions where all shots have been post-selected that were either erasures, or multiple logically distinct solutions were found. Here, forced gap cannot post-select any further.}
  \label{fig:idling_results}
\end{figure*}

\paragraph*{Results}

As mentioned, we instantiate our ``forced gap'' post-selection strategy with the Relay-BP decoder~\cite{muller2025improved}, primarily because it provides strong performance, while being fast and suitable for FPGA implementation~\cite{maurer2025real,maurya2025fpga}. While convergence is not guaranteed, we find in testing that Relay-BP converges often enough on the forced runs to make forced gap post-selection work well.

We test forced gap post-selection on the $72$ and $144$-qubit BB code memories~\cite{bravyi2024high}, and on surgery constructions for the latter from~\cite{yoder2025tourgrossmodularquantum}, all under the the circuit depolarizing noise model common to both those prior works. It is known that a careful choice of \textit{memory parameters} for the Relay-BP decoder can dramatically improve its error correction performance~\cite{muller2025improved}, and so for each circuit tested we optimise those parameters over an initial sweep for logical error rate (with no post-selection) before testing forced gap post-selection with those values. The choices of memory parameters, and other parameter choices for Relay-BP, are included in Appendix~\ref{sec:parameter_choices}. Otherwise, our testing uses the implementation of Relay-BP as in~\cite{trmue_relay}.

The performance of forced gap post-selection on idling circuits is shown in Figure~\ref{fig:idling_results}. In both cases, very few erasure shots were found, but for the $72$-qubit code, a noticeable number of non-erasure shots were found with  exactly $\mathrm{Gap} = 0$, which we always post-select for $T>0$. This explains the space between the (black) ``no post-selection'' line, and the left-hand end of the (blue) post-selection curve there. In both cases, once all shots with finite gap are post-selected, the curve remains flat thereafter; what remains at this point are shots with $\mathrm{Gap} = \infty$ in which the baseline run found a solution but all forced runs did not converge. We see that Forced Gap performs well at reducing logical error rate with a minimal amount of post-selection, but does not continue to get better performance at post-selection rates much above $1\%$ for these idling circuits. With that said, for many post-selection use cases, like resource generation protocols, a post-selection rate of greater than $1\%$ may not be of interest.

Next, we present results for LDPC surgery circuits on the $144$-qubit BB code. Here, the circuits we tested are constructed using the methodology of~\cite{yoder2025tourgrossmodularquantum}, with one exception. As explained in that paper, the surgery circuits proceed via an initialisation and merging of the ancilla system with the code, a certain number of rounds of syndrome extraction in the merged system, and finally a splitting of the ancilla system with the code. In principle, the optimal number of rounds of syndrome extraction in the merged system should be $d_{\text{Circ}}$, the circuit distance, to balance the effects of errors in spacetime ($d_{\text{Circ}} = 10$ for all circuits considered here). On the other hand, it is possible that a number of rounds of syndrome extraction slightly above, or slightly below $d_{\text{Circ}}$, could lead to an improved logical error rate \cite{cross2024improved}. After optimising the memory parameters for a given surgery operation, we then attempt to optimise the number of rounds of syndrome extraction to improve the logical error rate without post-selection. We find, however, that a number of rounds of syndrome extraction equal to the circuit distance is essentially optimal in all cases, with only very slight differences. We list the number of rounds of syndrome extraction used for each gadget with the corresponding figure.

\begin{figure*}[t]
  \centering

  \begin{subfigure}[t]{0.48\textwidth}
    \centering
    \includegraphics[width=\linewidth]{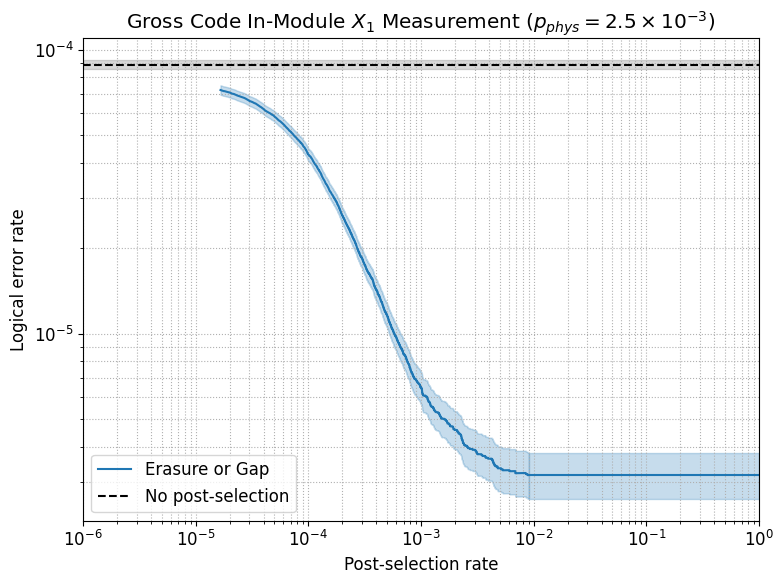}
    \caption{In-Module $X_1$ Measurement (10 Rounds)}
    \label{fig:postselection_72}
  \end{subfigure}
  \hfill
  \begin{subfigure}[t]{0.48\textwidth}
    \centering
    \includegraphics[width=\linewidth]{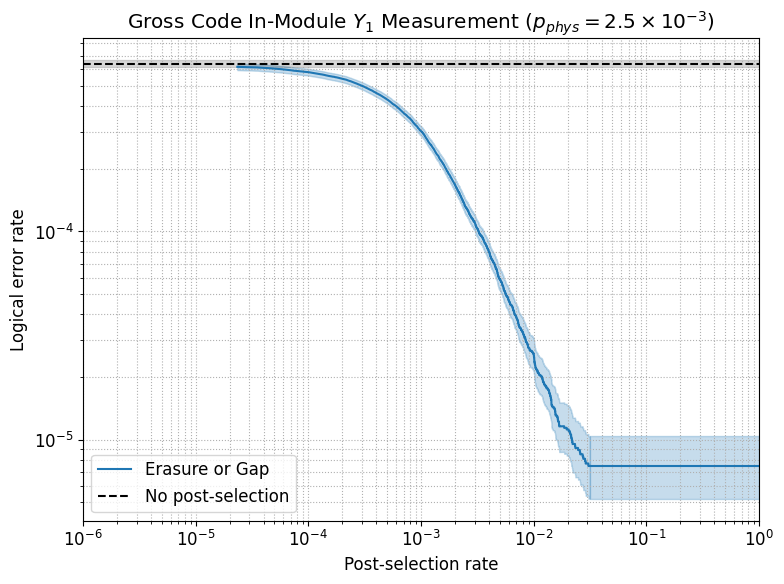}
    \caption{In-Module $Y_1$ Measurement (14 Rounds)}
  \end{subfigure}

  \vspace{0.6em}

  \begin{subfigure}[t]{0.48\textwidth}
    \centering
    \includegraphics[width=\linewidth]{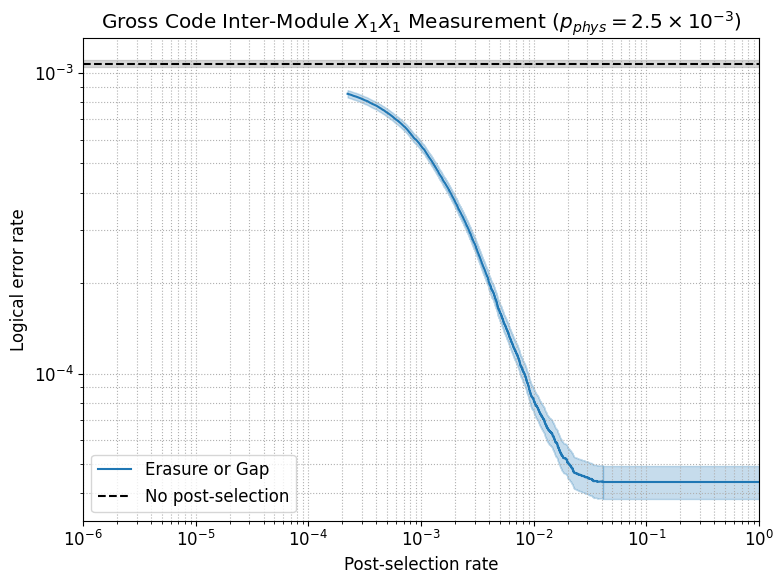}
    \caption{Inter-Module $X_1X_1$ Measurement (11 Rounds)}
  \end{subfigure}

  \caption{Logical error rates of various surgery bicycle instructions~\cite{yoder2025tourgrossmodularquantum} subject to post-selection under the forced gap strategy, all at physical noise $p = 2.5\times 10^{-3}$. We present the logical error rate for the entire circuit, without normalising by the number of rounds. The number of syndrome extraction rounds that we used in our simulations of each gadget are shown above. These were chosen by optimality of the logical error rate for regular error correction without post-selection. For the in-module $Y_1$ and inter-module $X_1X_1$ measurements, a higher number of rounds than $10$, the circuit distance, was seen to give an improvement, although the differences observed were minimal.}
  \label{fig:surgery_results}
\end{figure*}

In Figure~\ref{fig:surgery_results}, we present the results for the forced gap strategy on the in-module $X_1$, in-module $Y_1$, and inter-module $X_1X_1$ surgery instructions on the gross code~\cite{yoder2025tourgrossmodularquantum}, all at physical noise $p = 2.5\times 10^{-3}$. We see that the strategy continues to offer good performance on these logical instructions, with the post-selection curve flattening out around $1\%$ once again. For the in-module $X_1$ and inter-module $X_1X_1$ instructions, there is noticeable space between the left edges of the (blue) post-selection curve, and the (black) line showing the error rate with no post-selection. Interestingly, the reason for this space is different to the reason on the $72$-qubit code at idling (see Figure~\ref{fig:idling_results}). On the $72$-qubit code, there were very few erasure shots, but a larger number of non-erasure shots with exactly $\mathrm{Gap} = 0$. Here, there are a larger number of erasure shots, and very few non-erasure shots with exactly $\mathrm{Gap} = 0$.

\paragraph*{Relation to Prior Works}

The last few months have seen a notable amount of work done on post-selection strategies for quantum LDPC codes. First was~\cite{lee2025efficient}, based on clustering decoders, which inspired us to pursue this work. While preparing this work,~\cite{xie2026simple} appeared, and very recently also~\cite{gu2026scalable}. \cite{xie2026simple} offers a similar level of generality to ours, and is philosophically quite similar, where instead of ``forced runs'' (in which a logically distinct solution is forced of the decoder), the decoder is re-run with the physical likelihood of the initial correction suppressed. The primary focus of~\cite{gu2026scalable} is not to develop post-selection strategies, although it is shown that accurate and well-calibrated post-selection is possible using their neural decoder (see Figure 5 therein).

Both~\cite{lee2025efficient} and~\cite{xie2026simple} also test their strategies within the sliding-window framework, which is an important component of ultimately obtaining real-time and scalable decoding. One case not considered by the three works mentioned is testing the strategies on the logical instructions of high-rate codes, which is an important step towards their applicability to offline state preparation tasks like magic state distillation.

As for the performance of the various schemes, the only common point of comparison is the $72$-qubit code at physical noise $p = 1\times 10^{-3}$ (we test other circuits at higher physical noise). At the highest level of post-selection that it gets to, $1\%$, the forced gap strategy achieves a logical error rate per round of around $3.5 \times 10^{-8}$. By comparison,~\cite{lee2025efficient} achieves around $10^{-5}$, and~\cite{xie2026simple} around $1.5\times10^{-7}$.\footnote{We infer these numbers from Figure 4a of~\cite{lee2025efficient} and Figure 4 of~\cite{xie2026simple}, noting that both report logical error rate over $d=6$ rounds of syndrome extraction.}\footnote{For a further point of comparison, see Figure 5b of~\cite{gu2026scalable}, where post-selection on the $72$-qubit BB code is considered at physical error rates as low as $p = 3\times 10^{-3}$. There, a $1\%$ post-selection rate offers a logical error rate per round of about $1.8\times 10^{-5}$ (note that logical error rates in that paper are normalised by the number of logical qubits). However, this cannot be directly compared to ours and the other papers' values, which are all at physical noise $p = 10^{-3}$.} We see that forced gap post-selection can offer significant improvements. With this said, both of those other strategies can continue to offer improvements at higher rates of post-selection than $1\%$. In particular, whereas our strategy does not improve above its $1\%$ post-selection rate value of $3.5\times 10^{-8}$,~\cite{lee2025efficient} surpasses it at a post-selection rate of about $30\%$, and~\cite{xie2026simple} improves on it at a post-selection rate of about $5\%$, although note that the latter uses repeated runs of a heavy-duty BP-OSD decoder~\cite{panteleev2021degenerate}. One might draw the conclusion that forced gap post-selection is preferred in the regime of low post-selection rates (less than $1\%$), but other schemes are preferable in the higher post-selection rate regime. 

A direct comparison to the performance of the post-selection strategies in~\cite{gu2026scalable} (see Figure 5b therein) is hard, both because they consider higher physical error rates than $p = 10^{-3}$, and also because there they consider the (post-selected) logical error rate of one particular logical qubit, not the whole block.\footnote{Thanks to Andi Gu for clarifying this for us.}

A further work~\cite{sales2025experimental} also performed post-selection via comparative decoding, although this was across exponentially many decoding instances (rather than linearly many) using a maximum-likelihood decoder. This was feasible given the focus on a small, non-LDPC code for magic state distillation.

\paragraph*{Outlook}

In this work, we proposed a decoder-agnostic post-selection strategy for quantum LDPC codes with many logical observables. We instantiated and tested it with the Relay-BP decoder on bivariate bicycle codes and their instructions. In regimes of interest, our results show improved logical error rates relative to prior techniques, despite the relatively light-weight computation required. We believe these concepts are likely to provide utility for many situations where post-selection is available for quantum LDPC codes, especially offline resource state generation. One immediate application could be to the results of~\cite{xu2026distilling}, where magic states are distilled in BB codes, since it is shown that operational noise (i.e. logical measurements instructions) dominate for several key protocols, including the $15$-to-$1$ protocol~\cite{bravyi2005universal} on the gross code.

Future work could attempt to break the error floor that results from the forced runs failing to converge by introducing other decoding strategies for those runs specifically, such as following Relay-BP with ordered statistics decoding (OSD) \cite{panteleev2021degenerate} or by using decimation \cite{yao2024belief} to break up the high-weight row of the $H^{(i)}$ matrices that may be impeding convergence.

Another direction is to combine post-selection strategies like forced gap with rare-events analysis for fault-tolerant quantum systems \cite{bennett1976efficient,bravyi2013simulation,beverland2025fail} with the goal of accurately determining how post-selection scales to larger code distances and lower physical error rates without having to perform direct Monte Carlo simulation.

\end{multicols}

\section*{Acknowledgements}

The authors acknowledge helpful conversations with Michael E. Beverland, Lev S. Bishop, and Andrew W. Cross. The authors particularly thank Andi Gu for conversations on~\cite{gu2026scalable}.

AW acknowledges funding from NSF grant PHY-2325080. AW acknowledges funding from the MIT-IBM Watson AI Lab. AW acknowledges that this work is supported by the National Science Foundation under Cooperative Agreement PHY-2019786. This pre-print is assigned number MIT-CTP/6016.

This paper benefitted from computational resources. This research used resources of the National Energy Research Scientific Computing Center, a DOE Office of Science User Facility supported by the Office of Science of the U.S. Department of Energy under Contract No. DE-AC02-05CH11231 using NERSC award NERSC DDR-ERCAP0038585. This work used CPU resources at PSC Bridges-2~\cite{brown2021bridges} through allocation CIS260050 from the Advanced Cyberinfrastructure Coordination Ecosystem: Services \& Support (ACCESS) program~\cite{boerner2023access}, which is supported by U.S. National Science Foundation grants \#2138259, \#2138286, \#2138307, \#2137603, and \#2138296. We acknowledge the MIT Office of Research Computing and Data for providing high-performance computing resources that have contributed to the research results reported within this paper. We acknowledge the MIT SuperCloud~\cite{reuther2018interactive} and Lincoln Laboratory Supercomputing Center for providing HPC resources that have contributed to the research results reported within this paper. The computations in this paper were partially run on the FASRC Cannon cluster supported by the FAS Division of Science Research Computing Group at Harvard University. This work is supported by the National Science Foundation under Cooperative Agreement PHY-2019786 (The NSF AI Institute for Artificial Intelligence and Fundamental Interactions, http://iaifi.org/). This research was done in part using services provided by the OSG Consortium~\cite{osg_ospool_2006,ruth2007open,sfiligoi2009pilot,osg_osdf_2015}, which is supported by the National Science Foundation awards \#2030508 and \#2323298. AW is particularly grateful for the expertise and guidance of OSPool staff.
\clearpage
\printbibliography

\appendix

\section{Parameter Choices for the Relay-BP Decoder}\label{sec:parameter_choices}

The Relay-BP decoder~\cite{muller2025improved} is a recently-proposed heuristic decoder for quantum LDPC codes, shown to have good performance for correcting errors on surface codes, BB codes, their operations~\cite{yoder2025tourgrossmodularquantum}, and beyond~\cite{zhao2026towards}. Here, multiple ``legs'' of modified belief propagation (BP) instances are performed sequentially, with decoding information passed forward in the relay as described in~\cite{muller2025improved}. If a given leg converges to a candidate solution within a fixed number of iterations of BP, that candidate solution is recorded. Otherwise, the decoder moves onto the next leg. By default, the relay runs until a certain number of candidate solutions, $S$, has been found, or some maximum number of legs, $R$, has been run. Across all candidate solutions found, the most-likely of the candidate solutions is returned, or decoder failure if none were found.

Numerically benchmarking our strategies made use of the implementation of Relay-BP in~\cite{trmue_relay}, circuits constructed in~\cite{yoder2025tourgrossmodularquantum}, and Stim~\cite{gidney2021stim}. We also use $XYZ$ decoding throughout (see~\cite{muller2025improved}). We specify here the parameter choices that were made. The most important parameter choice is that of the $\gamma$ ``memory parameters''. These are two numbers $\gamma_{\min}$ and $\gamma_{\max}$, such that the memory strengths for each BP legs are drawn from a uniform distribution $[\gamma_{\min}, \gamma_{\max}]$. For example,~\cite{muller2025improved} uses $\gamma_{\min} = -0.24$ and $\gamma_{\max} = 0.66$ for the idling $144$-qubit code at physical noise $p = 3\times 10^{-3}$ noise. The performance of Relay-BP depends critically on a circuit-dependent choice of $\gamma_{\min}$ and $\gamma_{\max}$, as was found in~\cite{muller2025improved}, and as we found in our testing. Our choices for these memory parameters were made by an initial sweep, aiming to optimise the logical error rate at a post-selection rate of $0$ (regular error correction). 

We make a large choice of the parameter called $S$ in~\cite{muller2025improved}, and called $\mathtt{stop\_nconv}$ in the implementation~\cite{trmue_relay}. The average decoding time depends heavily on the choice of $\mathtt{stop\_nconv}$. When designing post-selection strategies, one is not so worried about the strategy keeping up with syndrome extraction, because one can reject decoding instances in the past. One cannot take $\mathtt{stop\_nconv}$ too large, though, because we do not want to wait too long for a post-selection decision, so as not to hold up the wider computation. Therefore, we take $\mathtt{stop\_nconv} = 100$. For context, the limit of real-time decoding, assuming the use of superconductors, running the decoder in an FPGA,\footnote{The use of an ASIC would speed the decoder up considerably.} and fixing our other parameters would be \textit{approximately} $\mathtt{stop\_nconv} = 10$~\cite{muller2025improved} for one run of Relay-BP.\footnote{The time to complete the forced runs depends how many are run in parallel, but if all are run in parallel, they would take a time roughly a quarter of the baseline run, given our other parameter choices.}

Another important parameter is called $R$ in~\cite{muller2025improved}, and called $\mathtt{num\_sets}$ in~\cite{trmue_relay}. Because most legs in forced runs do not converge, it is necessary to take a much smaller value of $\mathtt{num\_sets}$ in the forced runs than in the baseline run, since otherwise the runtime of the strategy would be impractical, even for testing. The parameters used are shown below, where we use variable names as in~\cite{trmue_relay}. The left-hand table shows the parameters common to all circuits, and the right-hand table shows the memory parameters used for each circuit:
\begin{table}[h]
\centering
\begin{minipage}{0.42\textwidth}
\centering
\begin{tabular}{l|l}
\texttt{stop\_nconv}              & 100 \\
\texttt{num\_sets} (baseline run)     & 1201 \\
\texttt{num\_sets} (forced runs)       & 25 \\
\texttt{gamma0}                   & 0.1 \\
\texttt{pre-iter}                 & 80 \\
\texttt{set-max-iter}             & 60 \\
\end{tabular}
\end{minipage}
\hfill
\begin{minipage}{0.52\textwidth}
\centering
\begin{tabular}{l|cc}
Circuit & $\gamma_{\min}$ & $\gamma_{\max}$ \\
\hline
$72$-Qubit Code Idling   & -0.19 & 0.26 \\
$144$-Qubit Code Idling & -0.24 & 0.66 \\
In-Module $X_1$       & -0.16   & 0.60  \\
In-Module $Y_1$       & -0.14   & 0.66  \\
Inter-Module $X_1X_1$       & -0.14   & 0.70  \\
\end{tabular}
\end{minipage}.
\end{table}

\end{document}